\documentclass[aps,pra,showpacs,twocolumn,amsmath,amssymb,groupedaddress]{revtex4-1}

\usepackage{graphicx}
\begin{document}

\title{ Solitons supported by complex PT symmetric Gaussian  potentials}

\author{Sumei Hu,$^{1,2}$ Xuekai Ma,$^1$ Daquan Lu,$^1$ Zhenjun
Yang,$^1$ Yizhou Zheng, $^1$ and Wei Hu$^{1,*}$}
\affiliation{$^1$Laboratory of Photonic Information Technology, South China Normal University,  Guangzhou  510631, P. R. China}
\affiliation{$^2$ Department of Physics, Guangdong University of Petrochemical Technology,  Maoming 525000, P. R. China}
\affiliation{Corresponding author: huwei@scnu.edu.cn}

\begin{abstract}
The existence and stability of fundamental, dipole, and tripole solitons in
Kerr nonlinear media with parity-time symmetric Gaussian complex
potentials are reported. Fundamental solitons are stable not only in deep
potentials but also in shallow potentials. Dipole and tripole solitons are
stable only in deep potentials, and tripole solitons are stable in deeper
potentials than for dipole solitons. The stable regions of solitons
increase with increasing potential depth. The power of solitons
increases with increasing  propagation constant or decreasing  modulation
depth of the potentials.
\end{abstract}

\maketitle 

\section{INTRODUCTION}
Quite recently  much attention has been paid to light propagation in
parity-time (PT) symmetric optical media in theory and experiment
\cite{PT2010-PRL,PT2011-oe,PT2009-PRL,PT2008-PRL,PT2010-Nature,PT2010-pra}.
This interest was motivated by various areas of physics, including quantum
field theory and mathematical physics
\cite{PT1998-prl,PT2002-prl,PT2003-czy,PT2008-pla,PT2007-pla,PT2007-jmp,PT2005-pla,PT2000-pla,PT2005-plb}.
Quantum mechanics requires that the spectrum of every physical observable
quantity is real, thus it must be Hermitian. However, Bender {\it et al}
pointed out that the non-Hermitian  Hamiltonian with PT symmetry can exhibit
entirely real spectrum \cite{PT1998-prl}.  Many people discussed the definition
of PT symmetric operator and its properties. A Hamiltonian with a complex PT
symmetric potential requires that the real  part of the potential must be an
even function of position, whereas the  imaginary part should be odd. It was
suggested that in optics the refractive index modulation combined with gain and
loss regions can  play a role in the complex PT symmetric potential \cite
{PT2005-ol}.

Spatial solitons have been studied since their first theoretical prediction
\cite{DSs1964-pl}. Recently, researchers have focused on composite multimode
solitons. Many composite multimode solitons are associated with dipole and
tripole solitons. In local Kerr-type media, fundamental solitons are stable,
whereas multimode solitons are unstable. Otherwise, multimode solitons have
been studied in nonlocal nonlinear media theoretically and experimentally
\cite{DSs2010-pra,DSs2007-pra,DSs2006-ol}. Many authors have paid much attention
to multimode solitons in optical lattices too
\cite{DSs2009-pra,DSs2006-oe,DSs2010-optik}.

In this paper, we find that dipole and tripole solitons can exist and be stable
in Kerr nonlinear media with PT symmetric Gaussian complex potentials. The
stabilities of fundamental, dipole, and tripole solitons are mainly determined by
their corresponding linear modes for low propagation constants or deep
potentials. Fundamental solitons are stable not only in deep potentials but
also in shallow potentials. But  dipole and tripole solitons are only stable
in deep potentials, and tripole solitons are stable in potentials deeper than
that for dipole solitons. The stable ranges of solitons increase with
increasing potential depth.

\section{MODEL}
We consider the (1+1)-dimensional evolution equation of beam propagation along
the longitudinal direction $z$ in  Kerr-nonlinear media with complex PT
potentials,
\begin{equation}\label{solution}
i\frac{\partial U}{\partial z}+\frac{\partial^2 U}{\partial x^2}
+T[V(x)+iW(x)]U+\sigma|U|^2U= 0.
\end{equation}
Here $U$ is the complex envelop of slowly varying fields, $x$ is the transverse
coordinate, and $z$ is the propagation distance. $V(x)$ and $W(x)$ are the real
and imaginary parts of the complex potentials, respectively, and $T$ is the
modulation depth. $\sigma=1$ represents the self-focusing propagation, and
$\sigma=0$ represents the linear situation. Complex PT symmetric Gaussian
potentials are assumed as
\begin{equation}\label{potentials}
V(x)=e^{-x^2},  \,\,\,\,  W(x)= W_0x e^{-x^2},
\end{equation}
where $W_0$ is the strength of the imaginary part. For complex PT symmetric
Gaussian potentials, all eigenvalues are real when the real part of the potentials
is stronger than the imaginary, i.e. $W_0<1.0$. Otherwise the eigenvalues are
mixed for $W_0\geq 1.0$ \cite{PT2001-Pla}.


We search for stationary linear modes and soliton solutions to Eq.
(\ref{solution}) in the form $U=f(x)\exp(i\lambda z)$, where $\lambda$ is the
propagation constant, and  $f(x)$ is the complex function satisfying the
equation
\begin{equation}\label{begin}
  \lambda f=\frac{\partial^2 f}{\partial x^2}
+T[V(x)+iW(x)]f+\sigma|f|^2 f.
\end{equation}
We numerically solve Eq. (\ref{begin}) for different parameters by
the modified square-operator method \cite{yang-2007}. To examine the
stability of solitons in PT Gaussian potentials, we search for
perturbed solutions to Eq. (\ref{solution}) in the form
\begin{equation}
U=e^{i\lambda z}\{f(x)+[g(x)-h(x)]e^{\delta
 z}+[g(x)+h(x)]^* e^{\delta^*z}\}, \nonumber
\end{equation}
where $g(x)\ll f(x)$ and $h(x)\ll f(x)$ are the perturbations,  and ``*'' means
complex conjugation. Substituting perturbed $U(x,z)$ into Eq. (\ref{solution}),
linearizing for $g(x)$ and $h(x)$,  the eigenvalue equations about $g(x)$ and
$h(x)$ can be derived
\begin{eqnarray}\label{increase}
 \delta g =&-i[\frac{d^2 h}{d x^2}-\lambda h +TVh-iTWg+2|f|^2 h  \nonumber \\
 &-\frac{1}{2}(f^2-f^{*2})g-\frac{1}{2}(f^2+f^{*2})h], \nonumber \\
 \delta h =&-i[\frac{d^2 g}{d x^2}-\lambda g +TVg-iTWh+2|f|^2
 g \nonumber \\&+\frac{1}{2}(f^2-f^{*2})h+\frac{1}{2}(f^2+f^{*2})g].
\end{eqnarray}
The growth rate $Re(\delta)$ can be obtained numerically by the
original-operator iteration method \cite{yang-2008}. If
$Re(\delta)>0$, solitons are unstable. Otherwise, they are stable.

\section{Fundamental Solitons}
\begin{figure}[htbp]
  \centering
  \includegraphics[width=7.5cm]{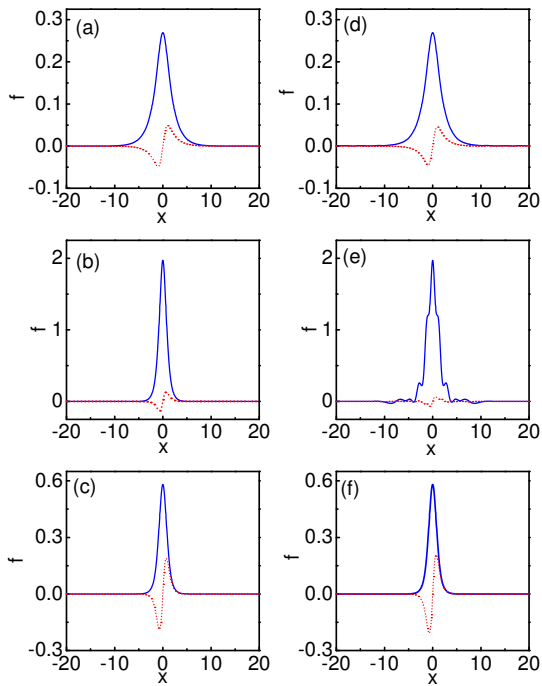}
  \caption{(Color online) Profiles of fundamental solitons with $W_0=0.1$ at
  (a) $T=1$, $\lambda=0.4$; (b)$T=1$, $\lambda=2.6$; and (c)$T=4$, $\lambda=2.6$.
  (d)-(f) are the linear modes corresponding to (a)-(c), respectively.
 Solid blue and dotted red lines represent  the real and imaginary
  parts of fields, and imaginary parts are multiplied by 10.} \label {Fig1}
\end{figure}

\begin{figure}[htbp]
  \centering
  \includegraphics[width=7.5cm]{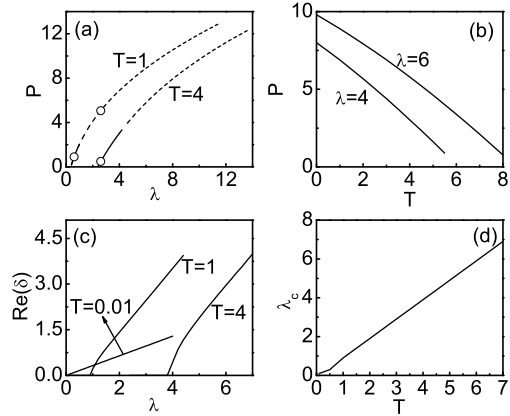}
  \caption{(a) Power $P$ versus propagation constant $\lambda$ with different modulated
depths $T$ for fundamental solitons. Solid lines represent stable range and
  dashed lines represent unstable range.
 (b Power $P$ versus modulated depth $T$ with different propagation constants
$\lambda$ for fundamental solitons.  (c)
 Perturbation growth rate versus propagation constant $\lambda$ with
 different $T$ values. (d) Critical propagation constant $\lambda_c$ versus modulated depth $T$.} \label {FigFSsP}
\end{figure}

We first investigate fundamental solitons in  PT symmetric Gaussian
potentials with $W_0=0.1$. Figures \ref{Fig1}(a)-\ref{Fig1}(c) show  fields
of fundamental solitons with different propagation constants and potential
depths, which correspond to the cases represented by circles in Fig.
\ref{FigFSsP}(a). We can see that all the real parts of fields are even
symmetric whereas the imaginary parts are odd symmetric.  With increasing 
propagation constant $\lambda$, the beam width narrows and the beam intensity
increases. With increasing potential depth, the beam intensity decreases but
the beam width changes little.

Figures \ref{Fig1}(d)-\ref{Fig1}(f) are the field distributions of linear modes
corresponding to Figs. \ref{Fig1}(a)-\ref{Fig1}(c), respectively. The field
distributions of linear modes and fundamental solitons are homologous for low
propagation constants [see Figs. \ref {Fig1}(a) and \ref {Fig1}(d)], or for
deep potentials [Figs. \ref {Fig1}(c) and  \ref {Fig1}(f)], but significantly
different for large propagation constants and shallow potentials [Figs. \ref
{Fig1}(b) and \ref {Fig1}(e)]. This phenomenon can be explained qualitatively
by Eq. (\ref{solution}). The nonlinear waveguide produced by the term $|U|^2
U$, along with the real part of the PT potential [$V(x)$], confines the expansion
of the beam induced by diffraction, and also suppresses the transverse energy flow
induced by the imaginary part of  the PT potential [$W(x)$]. Stationary linear
modes or solitons are obtained when all these effects are in balance. When the
propagation constant is small, the intensity of fundamental solitons and the
term  $|U|^2 U$ are small too [see Fig. \ref{FigFSsP}(a)]. The influence of
nonlinearity in Eq. (\ref{solution}) is weak, so the field distributions of
fundamental solitons are similar to those of corresponding linear modes. The
field distributions of linear modes and fundamental solitons are different when
the nonlinear term is comparable with the term $V$, i.e., for large propagation
constants and shallow potentials [Figs. \ref {Fig1}(b) and \ref {Fig1}(e)].

\begin{figure}[htbp]
  \centering
  \includegraphics[width=7.5cm]{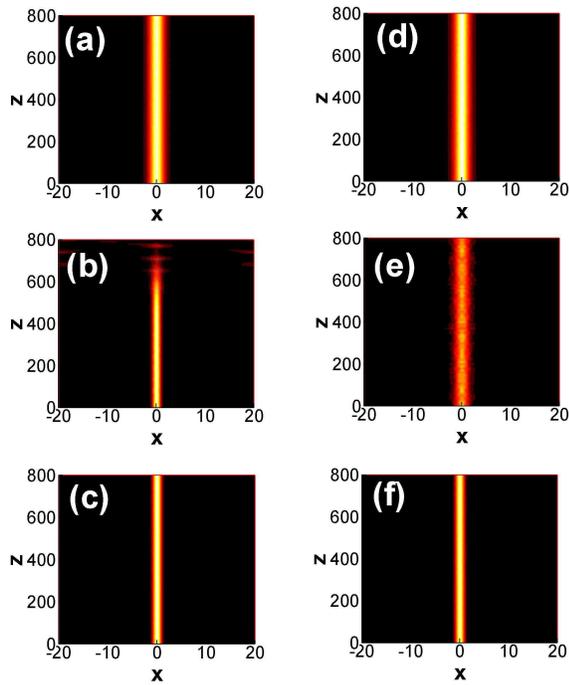}
  \caption{(Color online)
(a)-(c): Evolution of fundamental solitons corresponding to Figs.
\ref{Fig1}(a)-\ref{Fig1}(c), respectively. (d)-(f): Evolution of fundamental linear modes corresponding to Figs.
\ref{Fig1}(d)-\ref{Fig1}(f), respectively.} \label {Fig2}
\end{figure}

The power of solitons is defined as $P=\int_{-\infty}^{+\infty}|f(x)|^2 dx$.
Figure \ref{FigFSsP}(a) shows the power of solitons versus the propagation
constant $\lambda$ with different $T$ values, and Fig. \ref{FigFSsP}(b) shows the
power of solitons versus the potential depth $T$ with different $\lambda$ values. We
can see that the power of solitons increases with increasing $\lambda$ or
decreasing $T$. Figure \ref{FigFSsP}(c) shows the perturbation growth rate
versus the propagation constant $\lambda$ with different $T$ values. One can see that  the
stable range of fundamental solitons is $\lambda < \lambda_c$ for a fixed $T$,
where $\lambda_c$ is a critical propagation constant for soliton stability.
Figure \ref{FigFSsP}(d) shows that $\lambda_c$ is approximately proportional to
the modulated depth $T$. As $T$ approaches zero, $\lambda_c$ approaches zero
too, but the value of the growth rate $Re(\delta)$ decreases entirely [see the
curve of $Re(\delta)$ for $T=0.01$ in Fig. \ref{FigFSsP}(c)]. When $T=0$,
$Re(\delta)=0$ in the whole range, and solitons are always stable. This is consistent
with fundamental solitons in pure Kerr nonlinear media always being stable.

To confirm the results of the linear stability analysis, we simulate the
soliton propagation based on Eq. (\ref{solution}) with the input condition
$U(x,z=0)=f(x)[1+\epsilon \eta (x)]$ by the split-step Fourier method, where
$\eta(x)$ is a random function with a value of  between $0$ and $1$. $\epsilon$ is
a perturbation constant, which is 10 \% in our simulation. Figure \ref {Fig2}
shows the evolution of beams corresponding to those in Fig. \ref {Fig1}, which
is in agreement with the stability analysis in Fig. \ref {FigFSsP}. When
fundamental solitons are stable, the corresponding linear modes propagate with
no distortion [Figs. \ref{Fig2} (d) and \ref{Fig2}(f)]. This means that the
linear modes absorb the energy of the perturbation noise and maintain its mode
profiles. When fundamental solitons are unstable, the corresponding linear
modes propagate with random distortions [Fig. \ref {Fig2}(e)]. This indicates
that the linear modes and perturbation propagate independently without energy
exchange. According to Figs. \ref {Fig1} and  \ref {Fig2}, we can see that
the stability of fundamental solitons are  mainly decided by the complex
potentials for low propagation constants or deep potentials.

\begin{figure}[htbp]
  \centering
  \includegraphics[width=7.5cm]{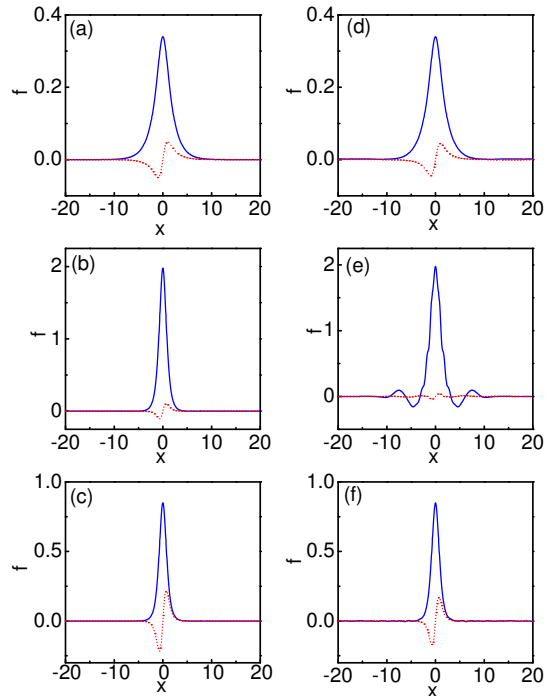}
  \caption{(Color online)
 Profiles of fundamental solitons with $W_0=0.8$ at (a) $T=1$, $\lambda=0.4$;
(b)$T=1$, $\lambda=2.6$; and (c)$T=4$, $\lambda=2.6$.
  (d)-(f) Linear modes corresponding to (a)-(c), respectively.
 Solid blue and dotted red lines represent  real and imaginary
  parts of fields.} \label {Fig7}
\end{figure}

\begin{figure}[htbp]
  \centering
  \includegraphics[width=7.5cm]{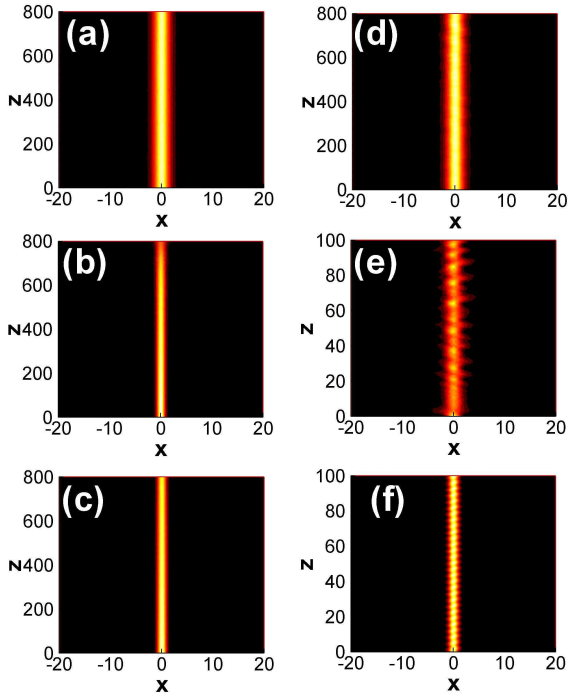}
  \caption{(Color online)
(a)-(c): Evolution of fundamental solitons corresponding to Figs.
\ref{Fig7}(a)-\ref{Fig7}(c), respectively. (d)-(f):  Evolution of  fundamental linear modes corresponding to Figs.
\ref{Fig7}(d)-\ref{Fig7}(f), respectively.} \label {Fig8}
  \end{figure}

We  also study fundamental solitons in the PT symmetric Gaussian potentials
with $W_0=0.8$. Figures \ref{Fig7}(a)-\ref{Fig7}(c) show the field
distributions of fundamental solitons with different propagation constants and
potential depths for $W_0=0.8$, whereas Figs. \ref{Fig7}(d)-\ref{Fig7}(f) are
the linear modes corresponding to them. We can see that the properties of
fundamental solitons for different $W_0$ are very similar, except the imaginary
parts of fields for $W_0=0.8$ are larger than those for $W_0=0.1$. Figure
\ref{Fig8} shows the beam evolutions corresponding to those in Fig. \ref{Fig7}.
We can see that fundamental solitons can propagate stably although $W_0$ is close
to the point of PT breaking [Figs. \ref{Fig8}(a) and \ref{Fig8}(c)].

\section{Dipole and Tripole Solitons}

\begin{figure}[htbp]
  \centering
  \includegraphics[width=7.5cm]{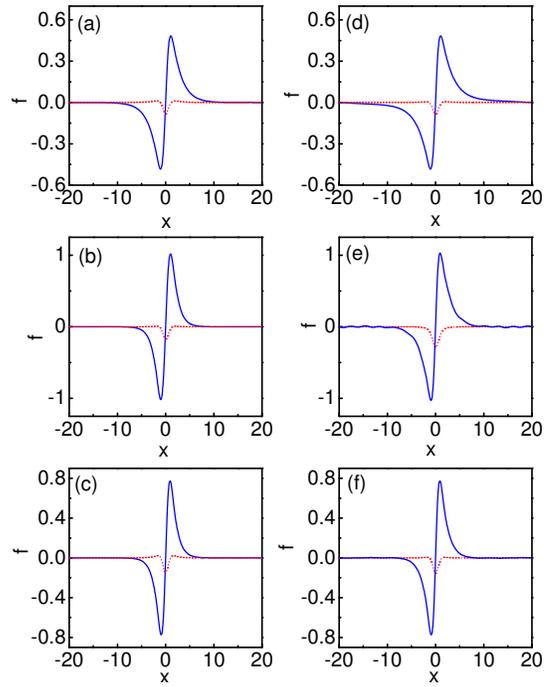}
  \caption{(color online)
  Profiles of dipole solitons with $W_0=0.1$ at (a) $T=4$, $\lambda=0.3$; (b)$T=4$, $\lambda=0.8$; and (c)$T=5$, $\lambda=0.8$.
  (d)-(f) Linear modes corresponding to (a)-(c), respectively.
 Solid blue and dotted red lines represent real and imaginary
  parts of fields, and imaginary parts are multiplied by 2.} \label {Fig3}
\end{figure}

\begin{figure}[htbp]
  \centering
  \includegraphics[width=7.5cm]{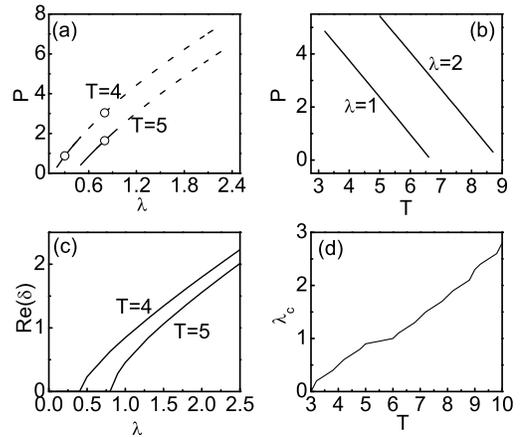}
  \caption{(a) Power $P$ versus propagation constant $\lambda$ with different
modulated depths $T$ for dipole solitons; Solid lines represent stable range and
dashed lines represent unstable range.
  (b) Power $P$ versus modulated depth $T$ with different propagation constants $\lambda$ for dipole solitons.
  (c) Perturbation growth rate versus propagation constant $\lambda$ with
 different $T$ values.
 (d) Critical propagation constant $\lambda_c$ versus modulated depth $T$.} \label {FigDSsP}
  \end{figure}

We now investigate dipole solitons in  PT symmetric Gaussian potentials with
$W_0=0.1$. Figures \ref{Fig3}(a)-\ref{Fig3}(c) show the field distributions of
dipole solitons, which correspond to the cases represented by circles
in Fig. \ref{FigDSsP}(a). Figures \ref{Fig3}(d)-\ref{Fig3}(f) are the linear
modes corresponding to Figs. \ref{Fig3}(a)-\ref{Fig3}(c). We can see that all
the real parts of the fields are odd symmetrical and the imaginary parts are
even symmetrical, which is converse to the situation for fundamental solitons. It is noteworthy  that all of the field distributions of linear modes and dipole solitons
are similar in Fig. \ref{Fig3}. Due to the deep potentials and small
propagation constants, the field distributions of solitons are decided mainly
by  PT symmetric Gaussian complex potentials.

The changes in the power versus $\lambda$ and $T$ for dipole solitons are shown
in Figs. \ref{FigDSsP}(a) and \ref{FigDSsP}(b), respectively. The power of
solitons increases with increasing  $\lambda$ or decreasing  $T$, which is
similar to the situation for fundamental solitons. Figure \ref{FigDSsP}(c) shows the perturbation
growth rate versus the propagation constant $\lambda$ for different $T$. Figure
\ref{FigDSsP}(d) shows the critical propagation constant $\lambda_c$ versus
the modulated depth $T$. We can see that dipole solitons exist stably only in 
deep potentials , i.e. $T\geq 3$, and the stable range increases with
increasing  modulation depth $T$.

Dipole solitons are always unstable for $T=0$, which corresponds to pure Kerr
nonlinear propagation. A dipole soliton can be considered  two solitons with a
$\pi$ phase difference, and a repulsive force exists between them.
However, we find that the dipole solitons are stable in  deep PT symmetric
Gaussian potentials. The reason is that the inherent repulsive interaction
between solitons can be effectively overcome by the real parts of the complex
PT symmetric potentials. This is the reason that dipole solitons are stable
only in deep potentials.

\begin{figure}[htbp]
  \centering
  \includegraphics[width=7.5cm]{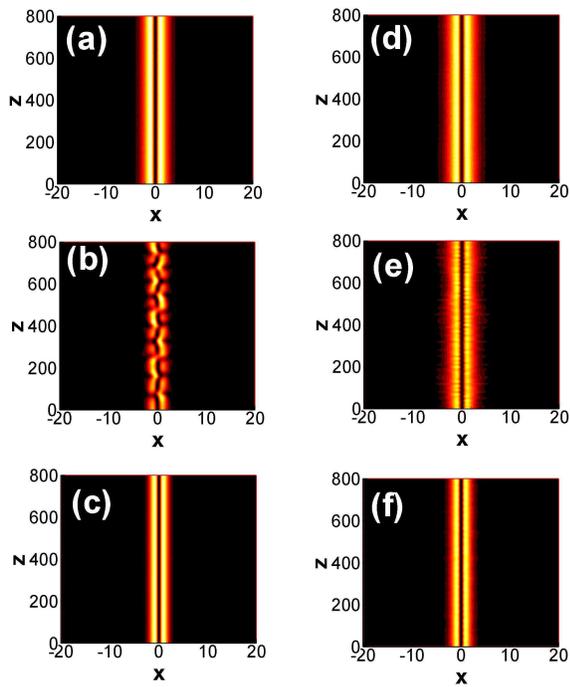}
  \caption{(Color online)
(a)-(c): Evolution of dipole solitons corresponding to Figs.
\ref{Fig3}(a)-\ref{Fig3}(c), respectively. (d)-
(f):  Evolution of dipole linear modes corresponding to Figs. \ref{Fig3}(d)-\ref{Fig3}(f), respectively.} \label {Fig4}
  \end{figure}

Figure \ref{Fig4} shows the beam evolutions corresponding to those in Fig. \ref
{Fig3}, which is agreement with the stability analysis in Fig. \ref{FigDSsP}.
The relations of the propagation between dipole solitons and the corresponding
linear modes are similar to those between fundamental solitons and their linear
modes.

\begin{figure}[htbp]
  \centering
  \includegraphics[width=7.5cm]{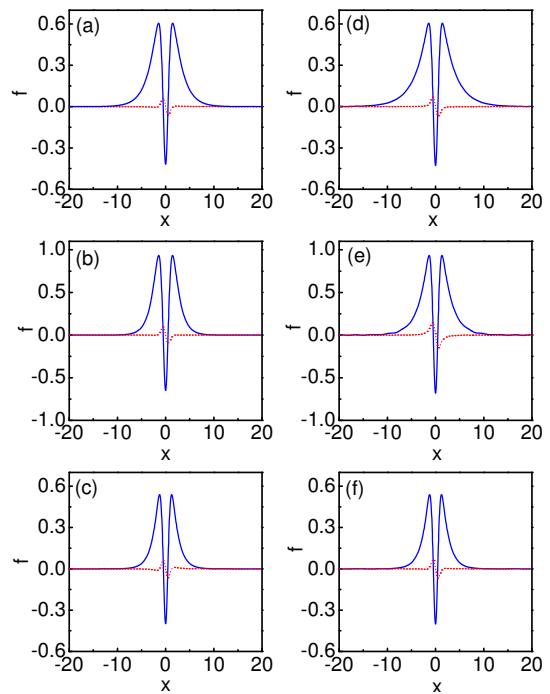}
  \caption{(Color online)
Profiles of tripole solitons with $W_0=0.1$ at (a) $T=10$, $\lambda=0.3$;
(b)$T=10$, $\lambda=0.6$; and (c)$T=12$, $\lambda=0.6$.
  (d)-(f) Linear modes corresponding to (a)-(c), respectively.
 Solid blue and dotted red lines represent  real and imaginary
  parts of fields.} \label {Fig5}
  \end{figure}
\begin{figure}[htbp]
  \centering
  \includegraphics[width=7.5cm]{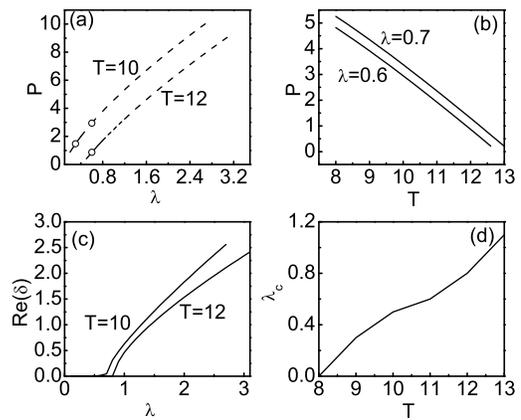}
  \caption{(a) Power $P$ versus propagation constant $\lambda$ with different modulated
depths $T$ for tripole solitons. Solid lines represent stable range and dashed
lines represent unstable range. (b) Power $P$ versus modulated depth $T$
with different propagation constants $\lambda$ for tripole solitons. (c) Perturbation growth rate versus propagation constant $\lambda$ with different
$T$ values. (d) Critical propagation constant $\lambda_c$ versus modulated depth
$T$.} \label {FigTSsP}
  \end{figure}

Finally, we study tripole solitons in PT-invariant Gaussian potentials with
$W_0=0.1$. Figure \ref{Fig5} shows the field distributions of tripole solitons
and their corresponding linear modes, which  correspond to the cases
represented by  circles in Fig. \ref{FigTSsP}(a). We can see that all the real
parts of the fields are even symmetrical and the imaginary parts are odd
symmetrical, which is similar to the fundamental solitons. Similar to dipole
solitons,  tripole solitons exist stably in deeper potentials , i.e. $T\geq
8$ [Fig. \ref {FigTSsP}(d)], and all of the field distributions of linear modes
and dipole solitons are similar.

The power of solitons increases with increasing  $\lambda$  and decreasing
$T$, which is similar to fundamental and dipole solitons, as shown in Figs.
\ref{FigTSsP}(a) and \ref{FigTSsP}(b). Figure \ref{FigTSsP}(c) shows the
perturbation growth rate versus propagation constant and Fig. \ref{FigTSsP}(d)
shows the critical propagation constant $\lambda_c$ versus modulated depth. We
can see that that tripole solitons are stable when $T \geq 8$, which is larger
than the value for dipole solitons. The reason is that a tripole soliton can be
considered  two pairs of out-of-phase solitons, and the repulsive force
between them is stronger than that for dipole solitons. Therefore, it needs
larger modulation depth $T$ to support tripole solitons than that for dipole
solitons.

\begin{figure}[htbp]
  \centering
  \includegraphics[width=7.5cm]{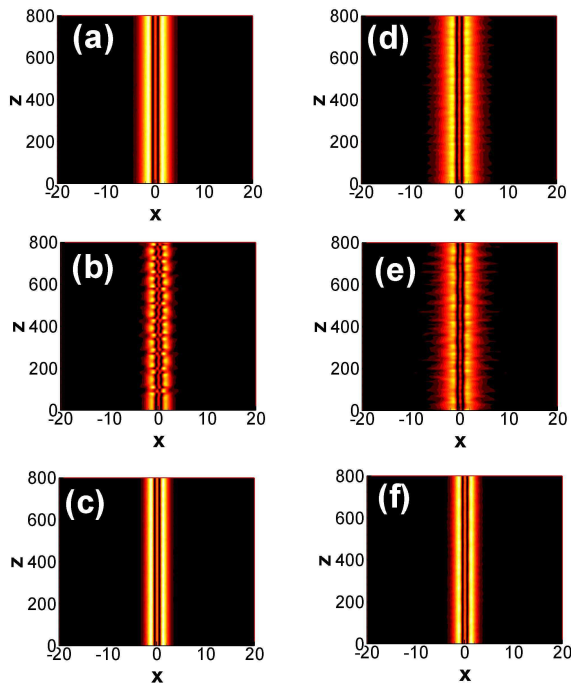}
  \caption{(Color online)
(a)-(c): Evolution of tripole solitons corresponding to Figs.
\ref{Fig5}(a)-\ref{Fig5}(c), respectively. (d)-
(f):   Evolution of tripole linear modes corresponding to Figs. \ref{Fig5}(d)
-\ref{Fig5}(f), respectively.} \label {Fig6}
\end{figure}

Figure \ref{Fig6} shows the beam evolutions corresponding to those in Fig.
\ref{Fig5}, which are in agreement with the stability analysis in Fig.
\ref{FigTSsP}.

\section{CONCLUSION}

In conclusion, we have reported the existence and stability of fundamental,
dipole, and tripole solitons supported by Gaussian PT symmetric complex
potentials. For monopole fundamental solitons, the waveguide effects from both
the nonlinearity and the real part of PT potentials are in balance with the
diffraction and the energy flows from the imaginary part of PT potentials. Thus
fundamental solitons are stable not only in deep potentials but also in shallow
potentials. For multipole solitons, a repulsive force exists between their
peaks, which needs a large modulation depth for counterbalance. Therefore
multipole solitons exist for a large modulation depth and relatively small
propagation constant. Our results may be extended to other PT symmetric optical
system, in which multipole solitons can exist.

Our model [Eqs. (\ref{solution}) and (\ref{potentials})] is given in
dimensionless form, where $x$ and $z$ are scaled to the potential width $a$ and
the diffraction length $L=2k_0n_0a^2$, respectively.  Here $k_0$ is the wave number
in vacuum and $n_0$ is the refraction index.  $a$ is defined as the half width
at the $1/e$ maximum of the real part of PT potentials [see Eq.
(\ref{potentials}) ], so the full width at half-maximum for $V(x)$ is about
$1.665a$ and the extreme of $W(x)$ is located at $\pm 0.707a$. The modulation
depth $T$ is scaled to the parameter $1/(2k_0^2 n_0a^2)$. Thus, for a typical
waveguide $a=10\mu m$ with the substrate  refraction index $n_0=3$, the
wavelength $\lambda_0=1.0\mu$m, and the diffraction length $L=3.77$mm.  Then
$T=1$ means that the maximum variation of refractive index is $4.22\times 10^{-5}$,
and $W=0.1$ means that the maximum gain/loss coefficient is about $0.53{\rm
cm}^{-1}$. For these physical parameters, it is feasible to realize 
multipole solitons in  synthetic PT symmetric systems. We hope that  the various
types of solitons may provide alternative methods in potential applications of
 synthetic PT symmetric systems.

\section*{ACKNOWLEDGMENTS} This research was supported by the
National Natural Science Foundation of China (Grant No. 10804033
and NO.11174090), the Program for Innovative Research No. 06CXTD005),
and the Specialized Research Fund for the Doctoral Program of Higher
Education (Grant No.200805740002).

\end{document}